\documentstyle[sprocl]{article}

\def\be{\begin{equation}}
\def\ee{\end{equation}}
\def\bea{\begin{eqnarray}}
\def\eea{\end{eqnarray}}
\begin{document}

\title{Pair creation of particles and black holes in external fields}

\author{\'OSCAR J. C. DIAS}

\address{ CENTRA, Departamento de F\'{\i}sica,
	      Instituto Superior T\'ecnico, 
Av. Rovisco Pais 1, 1096 Lisboa, Portugal\\E-mail: oscar@fisica.ist.utl.pt}

\maketitle\abstracts{
It is well known that massive black holes may form through the 
gravitational collapse of a massive astrophysical body. Less known is the
fact that a black hole can be produced by the quantum process of pair 
creation in external fields. These black holes may have a mass much lower 
than their astrophysical counterparts. This mass can be of the order of 
Planck mass so that quantum effects may be important.
This pair creation process can be investigated semiclassically using 
non-perturbative instanton methods, thus it may be used as a theoretical 
laboratory to obtain clues for a quantum gravity theory.
In this work, we review briefly the history of pair creation of particles and 
black holes in external fields. In order to present some features of the
euclidean instanton method which is used to calculate pair creation rates,
we study a simple model of a scalar field and propose an effective 
one-loop action for a two-dimensional
soliton pair creation problem. This action is built from the soliton 
field itself and the soliton charge is no longer treated as a topological 
charge but as a Noether charge. The results are also valid straightforwardly 
to the problem of pair creation rate of domain walls in dimensions D$\geq$3.
}

%%%%%%%%%%%%%%%%%%%%%%%%%%%%%%%%%%%%%%%%%%%%%%%%%
\noindent
{\bf 1. Black hole pair creation}
\vskip 3mm
%%%%%%%%%%%%%%%%%%%%%%%%%%%%%%%%%%%%%%%%%%%%%%%%%

Nowadays we have good observational evidence for black holes with a mass 
range between one solar mass and $10^{10}$ solar masses. These massive 
black holes have been produced through the gravitational collapse of 
massive astrophysical bodies.
One may be tempted to speculate on the possible existence of black holes
of much lower mass (of the order of Planck mass) for which quantum 
effects 
can be important. However, such black holes could not form from the 
collapse of normal baryonic matter because degeneracy pressure will 
support white dwarfs or neutron stars below the Chandrasekhar limiting 
mass.
Nevertheless, Planck size black holes may form through the tunneling 
quantum process of pair creation in external field. 

This kind of process was first proposed for  
electron-positron pair creation in the vacuum only due to the presence 
of an 
external electric field. Because of the vacuum quantum fluctuations,  
virtual 
electron-positron pairs are constantly being produced and annihilated.
These pairs can become real if they are pulled apart by an external 
electric field. The energy for the materialization and acceleration of
the pair comes 
from a decrease of the external electric field energy.
In the same way, a black hole pair can be created in the presence 
of an external field whenever the energy pumped into the system is 
enough in order to make the pair of virtual black holes real. The energy
for  black hole pair creation can be provided by a heat bath of gravitons\cite{H1,H2}, by a background 
electric field \cite{E1}$^{-}$\cite{E4}, by a background 
magnetic field \cite{M1,M2}, by a cosmic string \cite{S1}$^{-}$\cite{S3}, 
by a domain wall \cite{D1} or by a rapid cosmological expansion of the 
universe during the inflation era \cite{MannRoss}$^{-}$\cite{Volkov}.

Let us focus our attention on the process of black hole pair 
creation during inflation \cite{MannRoss}$^{-}$\cite{Volkov}. The 
inflationary era is not a good era to form black holes via gravitational 
collapse since matter is expanding away fast, rather than collapsing. 
However, this is a good era to create black holes through the 
quantum process of pair creation. The presence of large quantum 
fluctuations during inflation lead to strong gravitational perturbations 
and thus stimulates spontaneous black hole formation. Then, after the 
pair 
creation process, one has already  a force 
present which pulls the pair apart. Black holes will be separated by the 
rapid cosmological 
expansion due to the effective cosmological constant, $\Lambda_{\rm eff}$.
So, the cosmological expansion during the inflationary era prevents the 
black hole production via gravitational collapse, but provides the 
background needed for their quantum pair creation.

Using the instanton method, the pair creation rate 
for this process can be calculated. Pairs of black holes with a 
typical radius $r_{\rm BH}=1/ \Lambda_{\rm eff}$ are produced with a 
rate (in Planck units) given by 
$\Gamma \propto \exp{[-\pi /\Lambda_{\rm eff}]}$, so pair creation is
suppressed.
When $\Lambda_{\rm eff}\approx 1$ (early in inflation), the suppression 
is 
week and one can get a large number of black holes with a radius of      
order of Planck size. For smaller values of $\Lambda_{\rm eff}$ (later in
inflation), black holes are created with larger radius but their creation becomes
exponentially suppressed. 

After being pair created, as the inflaton field rolls down,  
$\Lambda_{\rm eff}$ decreases, and so the black hole grows slowly 
($r_{\rm BH}=1/ \Lambda_{\rm eff}$). 
However, the black hole also loses mass due to Hawking radiation and 
evaporates, so neutral black holes are highly suppressed after being 
pair created. The situation is different in the case of magnetically 
charged black holes which cannot evaporate, because either there 
are no magnetically charged particles they could radiate or they
are very massive. In spite of 
this, the pair creation of magnetic black holes is so small and 
they are so diluted
by the inflationary expansion that the probability to find one in our 
observable universe is extremely small.

We now briefly mention the other black hole creation processes. 
In the case of a heat bath of gravitons \cite{H1,H2} the creation 
process is not necessarily a pair creation. A single pair black hole can
pop out from the thermal bath. In fact,
due to statistical fluctuations, small black holes (with a temperature 
inversely proportional to the mass) can be produced in a thermal 
bath of gravitons. If the black hole's temperature is 
higher than the 
temperature of the background thermal bath then the black holes will 
evaporate by Hawking radiation. However, if the 
black hole's temperature is smaller than the temperature of the 
background thermal bath, the black hole will increase its mass 
by accretion of matter.

For electrically \cite{E1}$^{-}$\cite{E4} and magnetically  
\cite{M1,M2} charged black holes, the 
electromagnetic force  separates the recently created pair.

Black holes can also pair create in the background 
of a cosmic string \cite{S1}$^{-}$\cite{S3} or domain wall \cite{D1}. 
In these cases the force that keeps the 
black holes apart comes from the string and domain wall tensions.

In what follows we present a brief description of other studies on 
black pair production. The process of black hole pair creation has 
been studied in relation to black hole entropy \cite{O1}$^{-}$\cite{O4}, 
in de Sitter 
and AdS spacetimes \cite{O4}$^{-}$\cite{O10}, in instanton manifolds  
\cite{O11,O12}, in wormhole background \cite{O13}, in relation to Unruh 
effect \cite{O14}, within  the no-boundary proposal \cite{O15}, in 
an inflating brane-world \cite{O16}, and in primordial black hole 
setting \cite{O17}$^{-}$\cite{O19}.

%%%%%%%%%%%%%%%%%%%%%%%%%%%%%%%%%%%%%%%%%%%%%%%%%
\vskip 0.5cm

\noindent
{\bf 2. Particle pair creation in external fields reviewed}

\vskip 3mm
%%%%%%%%%%%%%%%%%%%%%%%%%%%%%%%%%%%%%%%%%%%%%%%%%

In order to better understand the black hole pair creation process 
we study 
now the particle pair creation process in external fields. In this 
section,
we present a brief historical review of this kind of process. Then, 
on the 
next section, we study a specific model of soliton pair creation in a 1+1 
dimensional scalar field theory. 

Klein \cite{Klein} has proposed the process of 
electron-positron pair creation in the vacuum due to the presence 
of an external electric field. This production process has been introduced
in order to solve Klein's paradox, which is related to the fact that 
the reflected plus the transmitted flux are greater than the incident 
flux 
when one considers the solution of Dirac's equation for an electron
entering into a region subjected to an external electric field.
Sauter \cite{Sauter} 
has shown that in order to materialize this pair, one has to 
have that the potential energy must satisfy  
$\Delta l V=eE\Delta l \geq 2 m c^2$ during approximately one Compton 
length,
$\Delta l \sim \hbar/mc$, so that the critical value for the electric 
field  
that one needs for the creation process is 
$E_{\rm cr} \sim 2.6 \times 10^{26} {\rm Vcm^{-1}}$. Heisenberg and Euler 
\cite{Heisenberg} have proposed, in the framework of electron-hole theory,
an one-loop effective lagrangian that accounts for the vacuum 
fluctuations 
effects and with it have calculated the electron-positron pair creation 
rate. Later, Schwinger \cite{Schwinger} has obtained the same result 
using 
a field theory approach by making use of his proper time method. 

Langer \cite{Langer}, in 1967, in his work about decay of metastable 
termodynamical states, has introduced the powerful euclidean instanton
method. As noticed by Stone \cite{Stone}, one can regard the external 
field as a false vacuum since its energy can be lowered by creating a 
pair of sufficiently separated particles. The  semiclassical  instanton 
method and Stone's
interpretation have been applied to several different studies namely:
Coleman and Callan \cite{Coleman1,Coleman2} 
have computed the bubble production rate that 
accompanies the cosmological phase transitions in a (3+1)D scalar field 
theory; Stone \cite{Stone}, Kiselev and Selivanov 
\cite{Kiselev1,Kiselev2} 
and Voloshin \cite{Voloshin} have calculated the soliton pair 
creation rate that accompanies the decay of a metastable vacuum on a 
(1+1)D scalar field theory; Affleck and Manton \cite{Affleck2} have 
studied monopole pair production in a weak external magnetic 
field; Affleck, Alvarez 
and Manton \cite{Affleck1} have worked on $e^+e^-$ boson pair production 
in a weak external electric field and finally the studies on black hole 
pair production in external fields \cite{H1}$^{-}$\cite{O19}.

For all these processes the 
instanton method can be used to compute
the pair creation rate, which is generally given by
$\Gamma=A\:\exp{[-(S^{\rm cl}_{\rm pair}-S^{\rm cl}_{\rm backg})]}$.
Here, $S^{\rm cl}_{\rm pair}$ is the classical action of the instanton 
mediating the pair creation,  $S^{\rm cl}_{\rm backg}$ is the classical 
action of the background field alone and pre-factor $A$ is the one-loop
contribution which includes the quantum corrections.

More recently, Miller and his collaborators have presented quite 
interesting 
experimental evidence for quantum pair creation of charged solitons 
in a condensed matter system \cite{Miller,Miller2}.

%%%%%%%%%%%%%%%%%%%%%%%%%%%%%%%%%%%%%%%%%%%%%%%%%
\vskip 0.5cm

\noindent
{\bf 3. The Instanton method. Effective one-loop action for pair 
creation of domain walls}

\vskip 3mm
%%%%%%%%%%%%%%%%%%%%%%%%%%%%%%%%%%%%%%%%%%%%%%%%%

Stone \cite{Stone} has studied the problem of a scalar field theory in
(1+1)D with a metastable vacuum, i.e., with a scalar potential $U$ that
has a false vacuum, $\phi_+$, and a true vacuum, $\phi_-$, separated 
by an energy density difference, $\epsilon$. Stone has noticed that the
decay process can be interpreted as the false vacuum decaying into the
true vacuum plus a creation of a soliton-antisoliton pair: $\phi_+
\rightarrow \phi_- +s +\bar{s}\:.$ The dynamics is governed by the
 action,
\begin{equation}
S[\phi(x,t)]=\int d^2x {\biggl [} \frac{1}{2}
\partial_{\mu}\phi \partial^{\mu}\phi + 
U(\phi) {\biggl ]}.
\label{1.1}
\end{equation}
We have proposed \cite{LemosDias} an euclidean effective 
one-loop 
action for Stone's problem which is built from the soliton 
field itself given by:
\begin{eqnarray} 
S^{\rm eff}_{\rm Euc}(\psi)=\int d^2x
{\biggl [} {\bigl |}(\partial_{\mu}-\frac{1}{2} \epsilon
\:\varepsilon_{\mu\nu}x_\nu)\psi{\bigl |}^2+m^2|\psi|^2 
{\biggr ]}\:.
\label{1.2}
\end{eqnarray} 
The action consists of the usual mass term
and a kinetic term in which the simple derivative of the 
soliton field is replaced by a kind of covariant 
derivative.  In this effective action the soliton 
charge is treated no longer as a topological charge but as a 
Noether charge. This procedure of working with an effective 
action for the soliton field itself is not new. Coleman 
\cite{Coleman3} has shown the equivalence between the 
Sine-Gordon model and the Thirring model. In this picture, 
the Sine-Gordon soliton is represented by the local Fermi
field of the Thirring model and there is an interchange
of Noether and topological charges. Montonen and Olive
\cite{Olive} have proposed an equivalent dual field theory
for the Prasad-Sommerfield monopole soliton \cite{Prasad} 
in which the fundamental monopole solitons fields play 
the role of heavy gauge particles, with the 
Prasad-Sommerfield topological magnetic charge being now 
a Noether charge. Manton \cite{Manton} has proposed an
effective action built from the soliton field  itself 
which reproduces the solitons' physical properties 
 of (1+1)D nonlinear scalar field theories that have 
symmetric potentials with degenerate minima. In the present 
problem one deals with (1+1)D scalar field theories  
which have a 
potential with non-degenerate minima, so our effective 
action is new since Manton was not dealing with 
the soliton pair production process.

Now, we can use the semiclassical instanton method to 
calculate the soliton-
antisoliton pair production rate per unit
time. Following Stone's interpretation this is equal to 
the decay rate per unit time 
of the false vacuum ($\hbar=c=1$): 
\begin{eqnarray} 
\Gamma=- 2\:{\rm Im}E_0\:. 
\end{eqnarray} 
One can see that it is so by considering the wavefunction 
associated with the metastable vacuum energy and analysing 
its probability evolution along the time.
The vacuum energy, $E_0$, is given by 
the euclidean path integral: 
\begin{eqnarray} 
e^{-E_0 T} = \lim_{T
\rightarrow \infty} \int [{\cal D}\psi]  [{\cal D}\psi^{\ast}]
e^{-S^{\rm eff}_{\rm Euc}(\psi,\psi^{\ast})}\:, 
\label{1.3}
\end{eqnarray} 
As it will be verified,
$E_0$ will receive a small imaginary contribution
from the negative-mode associated to the quantum 
fluctuations about the instanton (which 
stationarizies the action) and this fact is 
responsible for the decay.
After some calculations that make use of the
``Schwinger proper time integral" the creation rate
can be written as: 
\begin{eqnarray} 
\Gamma = \lim_{T
\rightarrow \infty}
\frac{1}{T}\:\frac{2}{m}\sqrt{\frac{2\pi}{{\cal{T}}_0}}{\rm Im}\int
[dx]e^{-S_{\rm Euc}[x_\mu(\tau)]} \:,
\label{1.4} 
\end{eqnarray}
where $S_{\rm Euc}=m\sqrt{\int_{0}^{1}
d\tau\dot{x}_\mu\dot{x}_\mu}+\frac{1}{2} \epsilon \oint
\varepsilon_{\mu\nu}x_\nu dx_\mu$ is now an effective action for a 
particle moving subjected to an external field in a (2+1) 
dimensional spacetime, with the Schwinger proper time $\tau$ 
playing the role of time coordinate. 

The classical solution of the equation of motion is called the
instanton, $x_\mu^{\rm cl}(\tau)=R(\cos2 \pi \tau, \sin 2 \pi
\tau)$, and represents a particle describing a loop of radius $R$ 
along the proper time.
The loop has a thin wall separating the true vacuum within from 
the false vacuum outside.
The euclidean action of the instanton is given by 
$S_0=S_{\rm Euc}[x_\mu^{\rm cl}(\tau)]=m 2 \pi R-\epsilon\pi R^2$.
The first term is the rest 
energy of the particle times the orbital length and the 
second term represents the interaction of the particle 
with the external scalar field.

Now, to include the quantum effects,
small fluctuations about the instanton are considered, i.e., the 
expansion $x_\mu(\tau)=x_\mu^{\rm cl}(\tau)+\eta_\mu(\tau)$. The 
euclidean action is expanded to second order so that the path integral
(\ref{1.4}) can be approximated by:
\begin{equation} 
\Gamma
\simeq \lim_{T \rightarrow \infty}
\frac{1}{T}\frac{2}{m}\sqrt{\frac{2\pi}{{\cal{T}}_0}}\:e^{-S_0}\:{\rm
Im} \int [d\eta(\tau)]\exp {\biggl[}-\frac{1}{2}\int d\tau
d\tau'\,\eta_\mu(\tau)\: M_{\mu\nu} \:\eta_\nu(\tau'){\biggr]}
\label{1.5} 
\end{equation} 
The path integral in equation (\ref{1.5}) is called the one-loop factor 
and
is given by 
${\cal N}{\bigl (}{\rm Det}M{\bigr )}^{-\frac{1}{2}}
={\cal N} \prod \,(\lambda_n)^{-\frac{1}{2}}$, where $\lambda_n$ are
the eigenvalues of $M_{\mu\nu}$, the second order variation operator 
of the action. Besides an infinite number of positive eigenvalues, 
one has two zero eingenvalues associated with the translation of the 
loop along  the $x_1$ and $x_2$ directions plus a zero eingenvalue
 associated with the 
translation along the proper time, $\tau$. 
There is also a single negative mode associated 
to the change of the loop radius. Note that it is this single 
negative eigenvalue, when one takes its square root, that is
 responsible for the imaginary contribution to the creation rate.

To overcome the problem of having a product of an infinite number 
of eigenvalues one has to compare our system with the background system 
without the pair created. In the productory, one omits the zero 
eigenvalues, but one has to 
introduce the normalization factor $\frac{||dx^{\rm cl}_\mu/d
\tau||}{||\eta^0_\mu||}\sqrt{\frac{1}{ 2\pi}}$  which is  
associated with the proper time eigenvalue.
In addition, associated with the negative eigenvalue one has to 
introduce a factor of $1/2$ which accounts for the loops that do 
expand. The other $1/2$ contracts (representing the annihilation of 
recently created pairs) and so does not contribute to the creation 
rate. One also has to introduce the spacetime volume 
factor $\int dx_2 \int d x_1=TL $, which represents the 
spacetime region where the instanton might be localized.

Finally, the soliton-antisoliton pair 
production rate per unit time and length is given 
by \cite{LemosDias}:
\begin{eqnarray}
 \Gamma/L = \frac{\epsilon}{2\pi}\:e^{-\frac{\pi
m^2}{\epsilon}}\:.  
\label{1.6} 
\end{eqnarray}
With our effective action (\ref{1.2}) we have recovered Stone's 
exponential factor $e^{-\frac{\pi m^2}{\epsilon}}$ \cite{Stone}
and the pre-exponential factor $A=\epsilon/2\pi$
of Kiselev, Selivanov and Voloshin
\cite{Kiselev1,Kiselev2,Voloshin}.

 One can make an analytical continuation of the euclidean time 
back to the Minkowskian time and 
obtain the solution in 2D Minkowski spacetime which tells us that
at t=0 the system makes a quantum 
jump and as a consequence of it a soliton-antisoliton 
pair materializes at $x=\pm R$. After 
the materialization, the 
soliton and antisoliton are accelerated, driving 
away from each other. The energy needed for this process
comes from the energy released when the false vacuum is
converted into 
true vacuum in the region  between the soliton pair. 

It is well known that a one-particle system in 2D can be transformed
straightforwardly to a thin line in 3D and a thin wall in 4D, where
now the mass $m$ of the soliton should be interpreted as a line
density and surface density, respectively.  (In fact a particle in
(1+1)D, as well as an infinite line in (2+1)D, can be considered as
walls as seen from within the intrinsic space dimension, justifying
the use of the name wall for any dimension).  Our calculations apply
directly to the domain wall pair creation problem in any dimension.

%%%%%%%%%%%%%%%%%%%%%%%%%%%%%%%%%%%
\vskip 0.5cm

\noindent
{\bf 4. Conclusions}
 
\vskip 3mm
%%%%%%%%%%%%%%%%%%%%%%%%%%%%%%%%%%%%%%%%%%%%%%%%%

In this work we have reviewed the possibility of producing
Planck size black hole pairs through the quantum tunneling 
process of pair creation. We have seen the principal features
of the semiclassical 
instanton method which is used to calculate particle pair creation 
rates in external fields. 
In particular, we have seen that the creation rate is given by the 
imaginary part of  
a path integral. The instanton is the classical solution
that stationarizes the euclidean action. Quantum corrections are 
included in the one-loop factor when one considers the quantum 
fluctuations around the instanton. An usual characteristic of the 
one-loop factor is the presence 
of: (i) an infinite product of eigenvalues; (ii) zero eigenvalues; 
(iii) a negative eigenvalue. It is this last one which is responsible 
for the creation rate.

\vskip .5cm
%%%%%%%%%%%%%%%%%%%%%%%%%%%%%%%
\section*{Acknowledgments} 
%%%%%%%%%%%%%%%%%%%%%%%%%%%%%%%

I would like to thank Jos\'e  Sande Lemos for suggesting the 
problem and many useful discussions. It is also a pleasure to thank 
Vitor Cardoso and Ana Mei Lin for all the encouragement.
This work was partially funded
by FCT through project ESO/PRO/1250/98. I 
also acknowledge financial support from the portuguese FCT 
through PRAXIS XXI program.

%%%%%%%%%%%%%%%%%%%%%%%%%%%%%%%%%%%%%%%%%%%%%%%%%

\end{document}